\documentstyle[emulateapj,apjfonts,psfig]{article}

\lefthead{J.~M.~Miller et al.}  
\righthead{Suzaku Observations of GX~339$-$4}

\received{Date}
\revised{Date}
\accepted{Date}

\journalid{vol}{date}
\articleid{1}{4}
\paperid{id}

\cpright{AAS}{1999}
\ccc{x}

\begin{document}

\title{Initial Measurements of Black Hole Spin in GX~339$-$4 from Suzaku Spectroscopy}

\author{J.~M.~Miller\altaffilmark{1},
        C.~S.~Reynolds\altaffilmark{2},
        A.~C.~Fabian\altaffilmark{3},
        E.~M.~Cackett\altaffilmark{1},
        G.~Miniutti\altaffilmark{3,4},
        J.~Raymond\altaffilmark{5},
        D.~Steeghs\altaffilmark{6},
	R.~Reis\altaffilmark{3},
	\& J.~Homan\altaffilmark{7}}

\altaffiltext{1}{Department of Astronomy, University of Michigan, 500
Church Street, Ann Arbor, MI 48109, jonmm@umich.edu}
\altaffiltext{2}{Department of Astronomy, The University of Maryland,
College Park, MD, 20742}
\altaffiltext{3}{Institute of Astronomy, University of Cambridge,
Madingley Road, Cambridge CB3 OHA, UK}
\altaffiltext{4}{Laboratoire APC, UMR 7164, 10 rue A. Domon et
L. Duquet, 75205 Paris, FR}
\altaffiltext{5}{Smithsonian Astrophysical Observatory, 60 Garden
Street, Cambridge MA, 02138}
\altaffiltext{6}{Department of Physics, University of Warwick,
Coventry CV4 7AL, UK}
\altaffiltext{7}{MIT Kavli Institute for Astrophysics and Space
Research, 70 Vassar Street, Cambridge, MA, 01239}


\authoremail{jonmm@umich.edu}

\label{firstpage}

\begin{abstract}
We report on a deep {\it Suzaku} observation of the stellar-mass black
hole GX~339$-$4 in outburst.  A clear, strong, relativistically-shaped
iron emission line from the inner accretion disk is observed.  The
broad-band disk reflection spectrum revealed is one of the most
sensitive yet obtained from an accreting black hole.  We fit the {\it
Suzaku} spectra with a physically--motivated disk reflection model,
blurred by a new relativistic line function in which the black hole
spin parameter is a variable.  This procedure yielded a black hole
spin parameter of $a = 0.89 \pm 0.04$.  Joint modeling of these {\it
Suzaku} spectra and prior {\it XMM-Newton} spectra obtained in two
different outburst phases yields a spin parameter of $a = 0.93 \pm
0.01$.  The degree of consistency between these results suggests that
disk reflection models allow for spin measurements that are not
strongly biased by scattering effects.  We suggest that the best value
of the black hole spin parameter is $a = 0.93 \pm 0.01~ {\rm
(statistical)} \pm 0.04~ {\rm (systematic)}$.  Although preliminary,
these results represent the first direct measurement of non-zero spin
in a stellar-mass black hole using relativistic line modeling.

\end{abstract}

\section{Introduction}
X-ray emission from accreting black holes probes the innermost
relativistic regime.  Emission lines produced in the inner accretion
disk are expected to bear the imprints of the strong Doppler shifts
and gravitational red-shifts natural to this region.  Advances in
X-ray spectroscopy have made it possible to exploit relativistic iron
lines as probes of black hole accretion and even measures of black
hole spin (for a review, see Miller 2007).  Recently, relativistic
line models have been developed in which black hole spin is a variable
parameter that can be constrained through spectral fitting
(e.g. Brenneman \& Reynolds 2007); this marks another advance.  {\it
Suzaku} is a major step forward for relativistic spectroscopy of
stellar-mass black holes: its CCD cameras retain a high livetime
fraction even at high flux levels, and its bandpass covers both the
relativistic iron line region and more subtle disk reflection
curvature expected at high energy.  The combination of new models and
the capabilities of {\it Suzaku} provide the opportunity to obtain
robust spin constraints in a number of stellar-mass black holes.

GX 339$-$4 is a recurrent black hole transient with a low-mass
companion star.  Its mass has been constrained to be greater than
$5.8~M_{\odot}$ (Hynes et al.\ 2004).  The distance to GX 339$-$4 is
likely to be greater than 6~kpc and may be as great as 15~kpc (Hynes
et al.\ 2004); a value of 8~kpc may be most likely (Zdziarski et al.\
2004).  The radio jet observed in GX~339$-$4 and its X-ray properties
argue for an inner disk inclination that may be as low as $15^{\circ}$
(Gallo et al.\ 2004, Miller et al.\ 2004a).  This inclination would
imply a high black hole mass, but the inner disk and binary system
need not have the same inclination.  

Prior observations of GX~339$-$4 with {\it Chandra} and {\it
XMM-Newton} have revealed lines suggestive of a black hole with a high
spin parameter (Miller et al.\ 2004a, 2004b, 2006).  Observations of
XTE J1650$-$500 with {\it XMM-Newton} and {\it BeppoSAX} have also
revealed skewed iron disk lines suggestive of high spin (Miller et
al.\ 2002; Miniutti, Fabian, \& Miller 2004; Rossi et al.\ 2005).
Spectroscopy of GRO~J1655$-$40 has sometimes revealed relativistic
iron lines suggestive of near-maximal black hole spin (Miller et al.\
2005; Diaz-Trigo et al.\ 2007).  In contrast to these sources, iron
lines observed in GRS~1915$+$105 does not require a high spin
parameter (Martocchia et al.\ 2002).

In late 2006, GX~339$-$4 entered a new outburst with a slow rise phase
(Swank et al.\ 2006).  In the spring of 2007, we triggered an approved
{\it Suzaku} TOO observation.  Based on {\it RXTE}/ASM monitoring, we
observed GX~339$-$4 in an ``intermediate'' state (for a review, see
Remillard \& McClintock 2006).  In the sections below, we detail how
we reduced and analyzed the spectra, and discuss the strengths and
weaknesses of our measurements and methods.

\section{Observations and Data Reduction}
We observed GX~339$-$4 with {\it Suzaku} on 2007 February 12 starting
at 05:33:31 (TT).  The nominal observation duration was 100~ksec, but
different exposures were obtained due to observing constraints and the
instrument modes chosen.  The XIS pointing was used.  In order to
prevent photon pile-up, the XIS units were operated using the 1/4
window mode using a 0.3~second burst option.  In general, XIS
on-source times of approximately 70~ksec were achieved.  This resulted
in deadtime-corrected net 3x3 editing mode exposures of 10.3~ksec,
9.3~ksec, and 16.9~ksec for the XIS0, XIS1, and XIS3 cameras,
respectively.  The HXD/PIN and HXD/GSO cameras were operated in their
default modes.  A net exposure time of 87.9~ksec was achieved using
the HXD/PIN. 

We reduced Version-2--processed data.  It is important to note that
the XIS burst modes are not yet calibrated as well as more standard
modes.  As reproducibility is even more important when non-standard
modes are used, we chose to use the XIS and HXD pipeline spectra.  For
the XIS, these products include both source and background spectra
from each camera; for the HXD/pin, the pipeling products include only
the source spectrum.  We generated XIS redistribution matrix files
(rmfs) and ancillary response files (arfs) using the tools
``xisrmfgen'' and ``xissimarfgen'' available in the HEASOFT version
6.4 reduction and analysis suite.  The ``xissimarfgen'' tool generates
a response using a simulation; we generated the recommended 400,000
photons in creating each ancillary response.  (XIS spectra over-sample
the true energy resolution of the instrument, as do the {\it
XMM-Newton} spectra treated in later sections.  Though rmf files
account for this, the reader should be aware that the
resolution is over-sampled.)  To create an HXD/PIN background file, we
simulated non-X-ray background and cosmic X-ray backgrounds in XSPEC
version 11.3 as per the instruction in the {\it Suzaku} ABC guide.
The standard PIN response file for XIS-nominal pointings was used.

The spectra from XIS0 and XIS1 matched each other closely, while the
spectrum from XIS3 has somewhat larger residuals (in the Si band, for
instance).  In order to focus on relativistic spectroscopy rather than
calibration uncertainties, we restricted our analysis to XIS0 and
XIS1.  For all plausible continuum models, these spectra show strong
residuals below 0.7~keV and above 10.0~keV; we therefore restricted
our XIS fitting range to the 0.7--10.0~keV band.  Similarly, the
HXD/PIN shows strong residuals below 12~keV.  The source is strong
throughout the HXD/PIN bandpass, so we considered the 12.0--70.0~keV
band when fitting this spectrum.  All spectral fits reported in this
work were made using XSPEC 11.3.  Errors reported in this work are
1$\sigma$ errors obtained using the ``error'' command.

\section{Analysis and Results}
The high sensitivity of our {\it Suzaku} spectra provides an
unprecedented chance to study a relativistic disk reflection spectrum,
and this is the focus of our analysis.  Especially given that the
absolute flux calibration in our modes is still being refined, and
given that contamination affects the effective area below 0.7~keV in
ways that could affect disk continuum spectroscopy, focusing only on the
residuals to fiducial continua is particularly pragmatic.  Indeed,
this is a general advantage of line and reflection spectroscopy over
disk continuum spectroscopy.  

Figure 1 shows the {\it Suzaku} spectra of GX 339$-$4 fit with a
simple model consisting of disk blackbody (``diskbb'') and power-law
components, modified by interstellar absorption (``phabs'').  The
continuum is the best--fit continuum for each individual camera (the
parameters were not linked during the fit).  A Gaussian emission line
was added at 2.26~keV to account for an instrumental response defect.
The 4.0-7.0~keV and 15.0-40.0~keV ranges were ignored when fitting the
data in order to show the disk reflection signatures clearly, and then
restored when forming the data/model ratio shown in Figure 1 (as per
Miniutti et al.\ 2007).

Fits to the continuum with this simple model reveals a 

\centerline{~\psfig{file=f1.ps,width=3.2in,angle=-90}~}
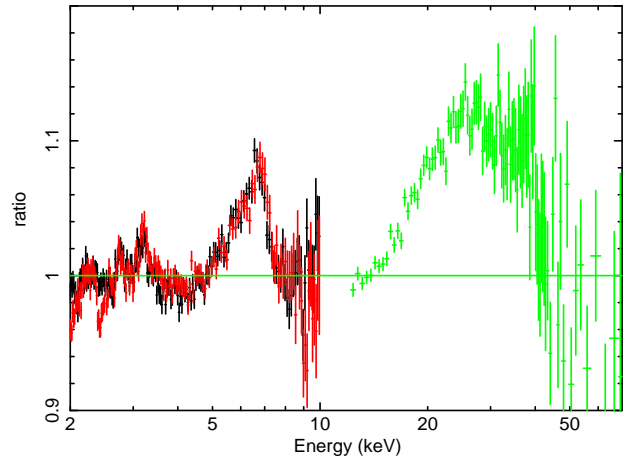
\figcaption[h]{\footnotesize The plot above shows the data/model ratio
obtained when the {\it Suzaku} spectra of GX~339$-$4 are fitted with a
phenomenological disk plus power-law model.  The 4.0--7.0~keV and
15.0--40.0~keV regions were ignored when fitting the model.  The
curvature at high energy is a clear signature of disk reflection.}
\medskip

\noindent  relativistic
iron emission line (Miller et al.\ 2004a, 2004b, 2006) and a Compton
back-scattering hump. The asymmetric shape of the broad line is
exactly that predicted by relativistic disk line models, and the
curvature of the Compton reflection hump is clearly revealed in the
HXD spectrum.  The broad-band relativistic disk spectrum is very
similar to that revealed in {\it Suzaku} spectra of the Seyfert-1 AGN
MCG-6-30-15 (Reeves et al.\ 2006, Miniutti et al.\ 2007).  It must be
noted that residuals remain in the 2--3~keV band.  These residuals are
not physical, but rather related to aspects of the instrument response
(e.g. Si and Au edges) that are not yet modeled correctly by the
response files.

Owing to the quality of the spectra line and reflection spectra, we
made fits with a physically-motivated model suited to constraining the
spin of the black hole in GX~339$-$4.  The model included a disk
blackbody component, a power-law component, and the constant density
ionized disk (CDID) reflection model (Ballantyne, Iwasawa, \& Fabian
2001).  This disk reflection model includes line emission, and was
convolved with the ``kerrdisk'' line function using the ``kerrconv''
model (Brenneman \& Reynolds 2006).  We have chosed to use the
``kerrdisk'' model because radii are parameterized in terms of the
innermost stable orbit for a given spin parameter; other line models
predict very similar profiles but differ in some details (see Dovciak,
Karas, \& Yaqoob 2004 and Beckwith \& Done 2004).  In all fits, the
outer reflection radius is assumed to be 400 times the innermost
stable circular orbit.  The CDID reflection model is particularly well
suited to high ionization regimes, and explicitly measures the inner
disk ionization parameter ($\xi = L/nr^{2}$).  The CDID model is an
angle-averaged model, and
does not include the disk inclination as a variable parameter.

This model provides a very good description of the spectra (see Figure
2 and Table 1), though the fit is not formally acceptable.  A high
inner disk ionization parameter of log($\xi$) = 4.1(2) is measured;
this is broadly consistent with prior results obtained in ``intermediate''
states (e.g. Miller et al.\ 2004b).  The most important result obtained
with this spectral model is the black hole spin parameter: $a =
0.89(4)$.

GX~339$-$4 is special in that excellent spectra have been obtained in
each of the spectral states where relativistic lines are expected.  We
next fit each of these spectra jointly to obtain the best possible
constraint on the black hole spin parameter.  

\centerline{~\psfig{file=f2.ps,width=3.2in,angle=-90}~}
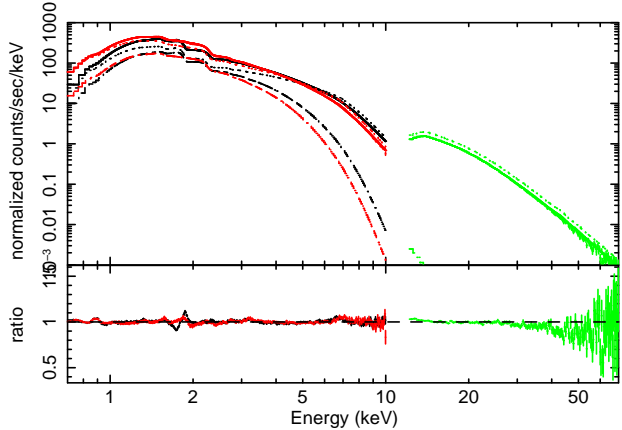
\figcaption[h]{\footnotesize The plot above shows the {\it Suzaku}
spectra of GX~339$-$4 fitted with a simple disk model and a constant
density ionized disk model (suited to highly ionized disks) convolved
with the ``kerrdisk'' line function.}
\medskip

\noindent This procedure also
ensures that state-specific changes (such as disk ionization) are not
driving any spin constraints.  Using the same model, we jointly fit
our {\it Suzaku} XIS spectra, the ``very high'' state spectrum
obtained with {\it XMM-Newton}/EPIC-pn camera, and the ``low/hard''
state spectra obtained with the {\it XMM-Newton} MOS1 and MOS2 spectra
in revolution 782.  (The HXD spectrum considered above was omitted as
its response is complex and drives up computation times.)
The same spectral and response files used in Miller et al.\ (2004a)
and Miller et al.\ (2006) were employed in this effort.  All spectra
were fit in the 0.7--10.0~keV band except the low/hard state spectra,
which were fit in the 0.7--9.0~keV band.  For simplicity, an
ionization parameter of ${\rm log}(\xi) = 4.0$ was assumed for the
intermediate state (see above), and a value of ${\rm log}(\xi) = 2.8$
was assumed for the low/hard state as per the results of modeling
described in Miller et al.\ (2006).  While each camera was again
allowed to have its own continuum parameters, the spin parameter and
inclination were determined by the joint fits.

This joint fit across states measures a spin parameter of $a =
0.93(1)$.  This is very similar to the result obtained using the
broad-band {\it Suzaku} spectra of the intermediate state alone, and
suggests that scattering in the disk atmosphere -- an effect expected
to change across states -- can accurately be accounted for with good
reflection models.  The very high state spectrum required an enhanced
inner line emissivity ($q = 6.0(1)$ within $6 GM/c^{2}$; $q = 3.0$ at
larger radii); this is common in the very high state (Miller et al.\
2002, 2004a; Diaz Trigo et al.\ 2007).  Once again, the overall fit is
not statistically acceptable in a formal sense, but this is driven by
narrow-band calibration problems.

\section{Discussion and Conclusions}
We have analyzed {\it Suzaku} spectra of the well-known recurrent
black hole transient GX~339$-$4 in outburst.  Motivated by the quality
of the broad-band spectra obtained, we fit the data with models
capable of measuring the black hole spin parameter.  Fits with a
relativistically blurred disk reflection model give a spin parameter
of $a = 0.89(4)$.  As a check on this result, we jointly fit the {\it
Suzaku} XIS spectra with CCD spectra in the ``very high'' and
``low/hard'' states obtained with {\it XMM-Newton}.  This procedure
yields a commensurate spin parameter of $a = 0.93(1)$.  We therefore
suggest that the best value of the spin parameter in GX~339$-$4 is $a
= 0.93 \pm 0.01 {\rm (statistical)} \pm 0.04 {\rm systematic}$, where
the systematic error aims to account for the error associated with
considering a single state.  Relativistic iron emission 

\centerline{~\psfig{file=f3.ps,width=3.2in,angle=-90}~}
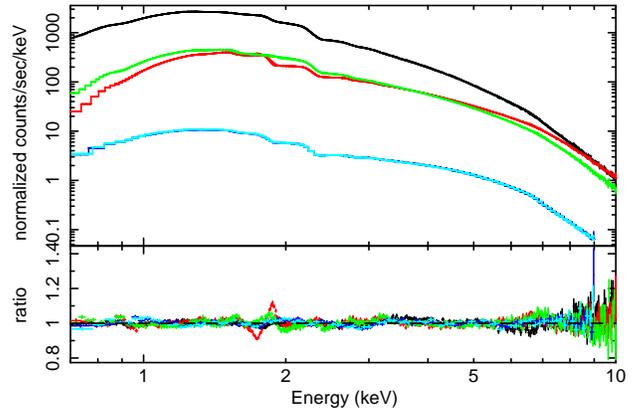
\figcaption[h]{\footnotesize The plot above shows the spectra of
GX~339$-$4 in the very high (black, {\it XMM-Newton}), intermeidate
(red and green, {\it Suzaku}), and low/hard (light and dark blue, {\it
XMM-Newton}) fitted with a simple disk model and a constant density
ionized disk model (suited to highly ionized disks) convolved with the
``kerrdisk'' line function.}
\medskip

\noindent lines
previously detected using {\it Chandra} and {\it XMM-Newton} also
suggested a similar spin parameter, but were not fit with
models wherein spin is a variable parameter (Miller et al.\ 2004a,
2004b).

We do not regard the apparent changes in line-of-sight column density
and inner disk radius in Table 1 as robust.  The changes are more
likely due in part to differences in flux calibration between {\it
XMM-Newton} and {\it Suzaku}.  The apparent changes may also reflect
real differences in disk ionization that falsely mimic radius
varations (Merloni, Fabian, \& Ross 2000).  Relativistic lines are
also likely to be affected by such changes, especially Compton
scattering in the disk atmosphere.  Disk reflection models attempt to
account for these effects.  Making joint fits to different states
allows us to obtain a measure of the spin parameter in a way that is
reasonably independent of changing scattering effects.

To achieve a maximal spin parameter, a black hole must accrete
approximately half of its mass (Volonteri et al.\ 2005).  It is
unlikely, then, that black holes in low-mass X-ray binary systems can
achieve maximal spin through accretion alone.  The short
lifespan of massive stars also makes it unlikely that black holes in
high-mass binaries can achieve a maximal spin parameter through
accretion.  Factoring in these constraints and the range of
black hole spins likely to be generated in supernova/GRB events,
Gammie, Shapiro, \& McKinney (2004) suggest that stellar-mass black
holes should have spins of $a \leq 0.95$.  The results we have
obtained are consistent with this prediction.

Accretion disks are necessarily an indirect measure of black hole
spin, and the accuracy of all such measurements depends on how strong
the contrast is between the inner edge of the accretion disk, and the
plunging region within the ISCO.  The most advanced 3-D general
relativistic MHD simulations of accretion disks are beginning to bear
on this problem, though the case of a standard thin accretion disk has
not yet been treated (Krolik \& Hawley 2002; Krolik \& Hirose 2004).
The case of a thin disk has been treated simply by Shafee, Narayan, \&
McClintock (2007); however, the most advanced and physically realistic
treatment is given by Reynolds \& Fabian (2008).  The latter work
makes use of sophisticated disk simulations and concludes that the
contrast between the disk and plunging region is relatively high and
that relativistic line diagnostics are robust.  This work partially
validates the implicit assumption of a geometrically-thin disk with
Keplerian orbits.

Although the relativistic modeling performed in this paper has
permitted a significant advance over many prior efforts to
characterize relativistic lines in stellar-mass black holes, our
results must still be regarded as preliminary.  In certain respects,
disk reflection models are like stellar atmosphere models: the physics
is known and the problems are tractable, but refinements can always be
made.  The CDID model we have used is excellent and well suited to
highly ionized disks (Ballantyne, Iwasawa, \& Fabian 2001), but
astrophysical disks have a vertical density structure.  Moreover,
averaging over viewing angle may act to wash-out important angle
dependencies.  At the time of writing, improved reflection models that
assume vertical structure according to hydrostatic equilibrium and a
hot thermal midplane spectrum are becoming available (Ross \& Fabian
2007; Reis et al.\ 2008).  All current reflection models assume a
single ionization zone; however, the fact that only the inner radii
contribute strongly to the red wing make this a robust simplification.

The use of relativistic lines to measure spin parameters in
stellar-mass black holes holds enormous promise and some specific
advantages over other techniques.  To obtain spin constraints using
the accretion disk continuum, the black hole mass, distance, and mass
accretion rate must be known.  The uncertainty on the mass can be as
high as 30\% (Remillard \& McClintock 2006), and the accretion rate
can be uncertain by a factor of a few.  As an observational rather
than empirical science, astronomy is poorly suited to absolute flux
(luminosity) measurements, and the difficulties with continuum
spectroscopy are merely a special case of the general problem.  Iron
lines have the advantage of harnessing dynamical imprints to estimate
the inner disk radius and black hole spin parameter.  A remaining
difficulty for relativistic iron line measurements is that the
geometry of the hard X-ray source illuminating the disk is unknown;
however, this is effectively encoded by the line emissivity index and
can be determined observationally.  

\vspace{0.1in}
We thank Charles Gammie, Laura Brenneman, and Randy Ross for helpful
discussions.  We thank the anonymous referee for helpful comments that
improved this paper.  This work has made use of the tools and services
available through HEASARC online service, which is operated by GSFC
for NASA.


\begin{table}[htb!]
\caption{Spectral Fit Parameters}
\begin{footnotesize}
\begin{center}
\begin{tabular}{lllllllllllll}
Spectra & ${\rm N}_{\rm H}$ & $kT$ & Norm & -- & $a$ & $i$ & q & log($\xi$) & $\Gamma$ & Norm. & Refl. Norm. & $\chi^{2}/\nu$ \\
~ & ($10^{21}~{\rm cm}^{2}$) & (keV) & ($10^{3}$) & ~ & ($cJ/GM^{2}$) & (degrees) & ~ & ~ & ~ & ~ & ($10^{-26}$) & ~ \\
\hline
Suzaku & 6.4(1) & 0.81(1) & 1.05(1) & ~ & $0.89(4)$ & $18(1)$ & 3.0(1) & 4.1(2) & 1.92(4) & 0.57(2) & 2.5(1) & 11860/5076 \\
\hline
XMM VH & 5.1(1) & 0.80(1) & 1.63(1) & ~ & $0.93(1)$ & $19(1)$ & 6.0(1) & 5.0(2) & 2.73(5) & 1.43(5) & 3.1(2) & 16190/7992 \\
Suzaku IS & 6.4(1) & 0.81(1) & 1.00(1) & ~ & -- & -- & 3.00(6) & 4.0 & 1.99(3) & 1.22(2) & 2.4(1) & -- \\
XMM LH & 4.3(1) & 0.44(1) & 0.47(1) & ~ & -- & -- & -- & 2.8 & 1.39(4) & 0.25(1) & 9.0(6) & -- \\
\tableline
\end{tabular}
\end{center} 
\tablecomments{The results of fitting models to the {\it Suzaku}
spectra of GX~339$-$4 and joint fits to different outburst states are
tabulated above.  Suzaku spectra from XIS0, XIS1, and the HXD/PIN were
fit jointly.  The model consists of a simple ``diskbb'' model plus a
constant-density ionized disk reflection model convolved with the
``kerrdisk'' function.  Relativistic line and reflection parameters
were linked while continuum parameters were allowed to vary between
cameras.  Parameters for XIS0 are quoted above in the ``Suzaku'' line;
parameters obtained for other cameras generally differ by less than
10\%.  Joint fits to {\it XMM-Newton} spectra in the very high (VH),
the {\it Suzaku} XIS spectra of the intermediate state (IS), and {\it
XMM-Newton} spectra of the low/hard (LH) state are presented below.
The spin parameter and inclination were linked across all spectra and
states, and ionization parameters were linked within states.  The
parameters for the EPIC-pn, XIS0, and MOS-1 camera are listed, and
once again the parameters for other cameras in each state generally
differ by less than 10\%.  The line/reflection emissivity index is
parameterized by ``q'', assuming $J(r) \propto r^{-q}$.  The
normalization of the reflection model is simly related to the number
of photons detected; it is a small number because distance dilution is
not included.  All errors quoted above are $1\sigma$ errors obtained
using the ``error'' command in XSPEC.}
\vspace{-1.0\baselineskip}
\end{footnotesize}
\end{table}

\end{document}